\documentstyle[aps,prl,multicol,epsf]{revtex}
\begin{document}
\draft
\title{Bi-layer Paired Quantum Hall States and 
Coulomb Drag}

\author{Yong Baek Kim$^1$, Chetan Nayak$^2$, Eugene Demler$^3$,
and Sankar Das Sarma$^4$}
\address{
$^1$ Department of Physics, The Ohio State University,
Columbus, OH 43210\\
$^2$ Department of Physics, University of California, 
Los Angeles, CA 90095\\
$^3$ Department of Physics, Harvard University, 
Cambridge, MA 02138\\
$^4$ Department of Physics, University of Maryland, 
College Park, MD 20742\\
}

\date{\today}
\maketitle

\begin{abstract}
We consider three states which are likely to
be realized in bi-layer quantum Hall systems
at total Landau level filling fraction ${\nu_T}=1$.
Two of these states may be understood as
paired states. One can occur as an
instability of a compressible state
in the large $d/l_B$ limit,
where $d$ and $l_B$ are the interlayer distance and magnetic
length.  The other is
the $(1,1,1)$ state, which is expected in the small $d/l_B$ limit.
We propose that the longitudinal
and Hall drag resistivities can be used
to distinguish these different quantum Hall states.
The drag resistivities may be computed in a straightforward manner
using the paired state formulation.
\end{abstract}

\pacs{PACS: }

\begin{multicols}{2}

\section{Introduction}

Bilayer quantum Hall systems at total Landau
level filling factor ${\nu_T}=1$ \cite{footnote_1}
allow for novel
interlayer coherent phases. These phases have attracted
a great deal of theoretical and experimental attention
\cite{Eisenstein97} over the last sixteen years, dating back
to a seminal paper by Halperin \cite{Halperin},
in which the multicomponent generalization of the
Laughlin wavefunction was first considered in a rather general context.
In particular, there is strong experimental
evidence \cite{Eisenstein97,eisenstein}
and a compelling theoretical basis \cite{Eisenstein97,Wen92}
to believe that a spin-polarized bilayer
$\nu={1\over 2}$ quantum Hall system would have
a novel spontaneous interlayer coherent
incompressible phase for small values of the interlayer
separation $d$, even in the absence of interlayer tunneling.
(We consider only the situation without any interlayer tunneling
in this paper -- our considerations also apply
to the physical situation with weak interlayer
tunneling. The situatiion with strong interlayer
tunneling is trivial by virtue of the tunneling-induced
symmetric-antisymmetric single-particle tunneling
gap which leads to the usual ${\nu_T}=1$ quantum Hall
state in the symmetric band.) In the limit $d\rightarrow\infty$,
however, one expects two decoupled layers, each
with $\nu={1\over 2}$, and hence no quantized Hall state.
The phase diagram for this incompressible to incompressible
quantum phase transition in $\nu={1\over 2}$ bilayer systems
has been studied extensively in the literature
\cite{Eisenstein97}, but we still do not have a complete
qualitative understanding of the detailed nature of this transition.
In particular, one does not know how
different kinds of incompressible (and compressible)
phases compete as system parameters (e.g. $d$)
are tuned, and how to distinguish among possible
competing incompressible phases. In this
paper, we revisit this issue by
arguing that, in principle, there are several interesting
and nontrivial quantum Hall phases in the
${\nu_T}=1$ bilayer system which could be systematically
probed via interlayer drag experiments carried out at various values
of the interlayer separation $d$.

Among the more interesting quantum Hall phases are the so-called
paired Hall states, which have been extensively
studied theoretically \cite{Halperin,Moore91,bonesteel}.
In these states, the composite fermions form an incompressible
superconducting paired state in which two electrons
composite fermions bind into effective Cooper pairs
which then condense into an incompressible ground state
analogous to the BCS state. The well-studied
Moore-Read Pfaffian state \cite{Moore91} is a
spin-polarized version of such a paired Hall
state for a single layer system.
Bilayer paired Hall states have been discussed
earlier in the literature in the context of
$\nu={1\over 2}$ (and $\nu={1\over 4}$ systems),
but no definitive idea has emerged regarding
their experimental observability or 
their relation to the more intensively studied
(and robust) $(1,1,1)$ state
\cite{Eisenstein97,Halperin,Wen92}.
The main purpose of the current paper is to critically discuss
the possible existence of a paired $\nu={1\over 2}$
bilayer Hall state which,
we argue, is distinguishable from
the better-studied $(1,1,1)$ incompressible
state (as well as from compressible states)
through interlayer drag experiments.
Given the great current interest \cite{eisenstein,demler}
in the phyiscs of $\nu={1\over 2}$ bilayer systems,
we believe that the results presented in this paper
could shed considerable light on the nature of the possible
quantum phase transitions in bilayer systems.

In sections II, III, and IV of this
paper, we consider the possible
bilayer $\nu={1\over 2}$ quantum Hall phases
in the parameter regimes $d\gg {\ell_B}$
(section II), $d> {\ell_B}$ (section III),
and $d\sim {\ell_B}$ (section IV), where
${\ell_B}$ is the magnetic length, which sets the
scale for intralayer correlations. We argue
that the likely ground states in these three
regimes are, respectively, compressible
(Fermi-liquid-like) states ($d\gg {\ell_B}$),
paired Hall states ($d> {\ell_B}$),
and $(1,1,1)$ states ($d\sim {\ell_B}$).
We conclude in section V with a critical
discussion of the various drag resistivities
which we argue can, in principle,
distinguish among these phases and could be used to
study bilayer quantum phase transitions.

\section{$\lowercase{d}\gg \ell_B$: Compressible State}

Let us consider a bi-layer system in which each layer has
filling factor $\nu=1/2$ in the limit that the layer separation
$d$ is much larger than the typical interparticle spacing,
which is of the order of the magnetic length $\ell_B$.
As a starting point, we will model the system
by two almost independent Fermi-liquid-like 
compressible states, one in each layer.
There are two alternative and complementary
descriptions of compressible states at even-denominator
filling fractions. We will briefly recapitulate
some features of both, as they will inform
the following discussion of paired states.

One description of a Fermi-liquid-like
compressible state at $\nu=1/2$ is based on the
lowest Landau level wavefunction \cite{rezayi}:
\begin{equation}
\label{eqn:RR_wavefcn}
\Psi_{1/2} (\{z_i\})
= {\cal P}_{LLL}\: {\rm Det} M
\prod_{i<j} (z_i  - z_j )^2 \ , 
\end{equation}
where $M$ has the matrix elements 
$M_{ij} = e^{i {\bf k}_i \cdot
{\bf r}_j}$. Here, ${\bf r}_i$
is the position of electron $i$. The  
${\bf k}_i$s are parameters which
are chosen so that 
the total energy of the system is minimized.
We will discuss their interpretation below.
${\cal P}_{LLL}$ projects states into the lowest Landau level;
it has the following action \cite{rezayi,girvin}
\begin{equation}
{\cal P}_{LLL} e^{i {\bf k}_i \cdot
{\bf r}_j} {\cal P}_{LLL} = 
e^{i {\bf k}_i \cdot
{\bf R}_j} \ , 
\end{equation} 
where ${\bf R}_i$ are the guiding center coordinates of the
electrons.

The corresponding wavefunction of the double-layer system
at ${\nu_T}=1$ can be written as
\begin{equation}
\Psi_{\rm compresible} 
= \Psi_{1/2} (\{z^{\uparrow}_i\}) 
\Psi_{1/2} (\{z^{\downarrow}_i\}) \ ,
\end{equation}
where $\alpha = \uparrow, \downarrow$ label the
two layers.

The lowest Landau level constraint and Fermi statistics
displace the electrons from the
correlation holes -- i.e. zeroes of
the wavefunction or, equivalently, vortices 
represented by the $\prod_{i<j} (z_i-z_j)^2$
factor in the wavefunction.
The ${\rm Det} M$
factor is necessary to ensure Fermi statistics;
it is a displacement operator \cite{read94} because
${\bf k}_i \cdot {\bf r}_j$ acts as
${\bar k}_i z_j + k_i {\partial \over \partial z_j}$ in the
lowest Landau level. 
Thus the composite fermion made of an electron and
two correlation
holes has a dipolar structure.
Using these ideas, the system
can be described as a collection of dipolar `composite 
fermions' in which each dipolar fermion consists of an electron
and the corresponding correlation holes
\cite{read94,shankar,stern,read98}.
To each composite fermion we assign a
${\bf k}_i$ which is equal in magnitude
(in units of $\ell_B$)
and perpendicular to the displacement
between the electron and its correlation holes.
By Fermi statistics, the ${\bf k}_i$s 
must be distinct. The structure of a dipolar composite fermion
is given schematically in Fig.1. The energy of a composite
fermion increases with ${\bf k}_i$,
so the ground state is a filled Fermi
sea in ${\bf k}$-space.
In the long wavelength limit, 
the total energy of these dipolar composite fermions can be 
approximately written as $\sum_i k^2_i / 2m^*$ with
effective mass $m^*$ which is
determined by the interaction 
potential \cite{shankar,read98}.

An alternative formulation springs from
the observation that an electron may be represented
by a `composite fermion' together with
a Chern-Simons gauge field which attaches two fictitious
flux quanta to each fermion \cite{jain,CS,HLR}.
This representation is mathematically equivalent
to the original one, but it naturally suggests
a new approximation. 
The composite fermions see zero average magnetic field due to the 
cancellation between the external magnetic field and
the average fictitious magnetic field coming from the
fictitious flux quanta. Consequently, the system
can be described as an almost free (composite)
fermion system in zero
effective magnetic field. In this approach, the fictitious flux
quanta are introduced to represent the phase winding of the
electron wavefunction around correlation holes associated
with the positions of other electrons.
The ${\rm Det} [e^{i {\bf k}_i \cdot {\bf R}_j}]$ factor
in (\ref{eqn:RR_wavefcn}) can
be interpreted as the wavefunction of the almost free
composite fermion system. \cite{rezayi,HLR} 
Then ${\bf k}_i$s can be
regarded as the ``kinetic momenta'' of
the composite fermions.
In the long-wavelength, low-frequency limit,
the two formulations are equivalent.

\begin{figure}    
\vspace{-0.5truecm}
\center    
\centerline{\epsfysize=1.2in    
\epsfbox{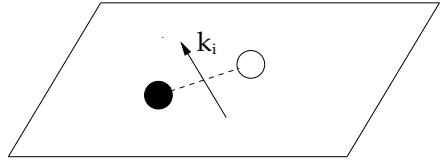}}
\vspace{-1.0truecm}    
\begin{minipage}[t]{8.1cm}    
\caption{Schematic picture of the dipolar composite
fermion. The black and white dots represent the 
electron and vortex respectively. The ``wavevector''
${\bf k}_i$ is perpendicular to the dipole moment.
}  
\label{fig1}   
\end{minipage}    
\end{figure}

\section{$\lowercase{d}>\ell_B$: Pairing}

We will now implement the latter formulation
in a double-layer system, and show
that the compressible state
has a strong pairing instability \cite{bonesteel}.
We introduce two composite fermion
fields $\psi_\alpha$ and two Chern-Simons
gauge fields ${\bf a}_\alpha$, $\alpha=\uparrow,\downarrow$.
The Hamiltonian is
\begin{eqnarray}
H &=& H_0 + H_I  \ , \cr
H_0 &=& \int d^2 r \sum_{\alpha = \uparrow, \downarrow} 
{1 \over 2m^*} \psi^{\dagger}_{\alpha} 
(\nabla - {\bf a}_{\alpha})^2 \psi_{\alpha} \ , \cr  
H_I &=& \int d^2 r \int d^2 r' \ 
\delta \rho_{\alpha} ({\bf r}) 
V_{\alpha \beta} ({\bf r}-{\bf r'})
\delta \rho_{\beta} ({\bf r'})   
\end{eqnarray}
with the constraints $\nabla \times {\bf a}_{\alpha}
= 2 \pi {\tilde \phi} \delta \rho_{\alpha} ({\bf r})$.
${\tilde \phi} = 2$ if the filling factor of each layer is $\nu=1/2$.
Here $\delta \rho ({\bf r}) = \rho ({\bf r})-{\bar \rho}$
is the density disturbance measured
from the average value ${\bar \rho}$.
The interaction potential is given by
$V_{\uparrow \uparrow} = V_{\downarrow \downarrow} = e^2/\varepsilon r$
and $V_{\uparrow \downarrow} = V_{\downarrow \uparrow} = 
e^2 / \varepsilon \sqrt{r^2 + d^2}$
where $\varepsilon$ is the dielectric
constant.

It is convenient to change the gauge field variables to
${\bf a}_{\pm} = {1 \over 2} 
({\bf a}_{\uparrow} \pm {\bf a}_{\downarrow})$.
Then $H_0$ can be rewritten as
\begin{eqnarray}
\label{eqn:H_0}
H_0 &=& \int d^2 r \left [
{1 \over 2m^*} \psi^{\dagger}_{\uparrow} 
(\nabla - {\bf a}_+ - {\bf a}_-)^2 \psi_{\uparrow} \right . \cr
&&+ \left . {1 \over 2m^*} \psi^{\dagger}_{\downarrow} 
(\nabla - {\bf a}_+ + {\bf a}_-)^2 \psi_{\downarrow}
\right ]
\end{eqnarray}   
with $\nabla \times {\bf a}_{\pm}
= \pi {\tilde \phi} [\delta \rho_{\uparrow}({\bf r})
\pm \delta \rho_{\downarrow}({\bf r})]$.
From (\ref{eqn:H_0}), we see that $\psi_{\uparrow}$
and $\psi_{\downarrow}$ have the same
gauge ``charges'' for ${\bf a}_{+}$, but opposite
gauge ``charges'' for ${\bf a}_{-}$. 
As a result, there will be an attractive interaction between the 
composite fermions in different layers via ${\bf a}_-$ and 
a repulsive interaction via ${\bf a}_+$. Composite fermions
in the same layer have repulsive interactions.
As a result of Coulomb intractions,
the attractive interaction mediated by the ${\bf a}_-$
gauge field dominates in the low energy limit and 
there exists a pairing instability between the composite
fermions in different layers.
This result can be understood in physical terms as follows.
The ${\bf a}_-$ and ${\bf a}_+$ fields represent antisymmetric 
and symmetric density fluctuations. In the presence of Coulomb
interactions, symmetric density fluctuations are highly 
suppressed, but antisymmetric density fluctuations can still
be large. As a result, the dynamic density fluctuations 
in the antisymmetric channel becomes more important in the
low energy limit and lead to a pairing instability.

This pairing instability has a natural
explanation in the dipolar composite fermion
picture \cite{read94,shankar,stern,read98}.
Let us take a dipolar composite fermion in layer 
$\uparrow$ with
wavevector ${\bf k}^{\uparrow} = {\bf k}_F$ and a dipolar
composite fermion in layer $\downarrow$ with wavevector
${\bf k}^{\downarrow}=-{\bf k}_F$.
As seen in Fig.2, this configuration can lower the interlayer
Coulomb energy because the electron in layer
$\uparrow$ and the vortex in layer $\downarrow$ can sit
on top of each other and vice versa.

\begin{figure}    
\center    
\centerline{\epsfysize=1.4in    
\epsfbox{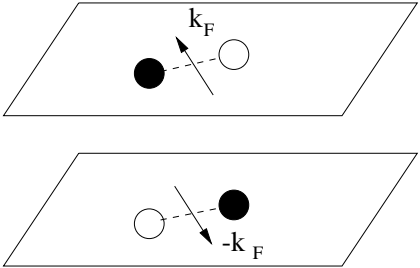}}
\begin{minipage}[t]{8.1cm}    
\caption{Pairing of the 
``intralayer'' composite fermions.
}  
\label{fig2}   
\end{minipage}    
\end{figure}    

This analysis predicts that, at least in principle,
the Fermi-liquid-like compressible state is
always unstable to pairing. In practice,
the pairing gap will be small in the limit
$d\gg \ell_B$ and easily destroyed by disorder.
As a result, we expect the pairing instability
discussed above to be relevant for $d/\ell_B$
not too small. When $d/\ell_B$ become small, on the
other hand, the starting point of two decoupled compressible states
is no longer sensible, and we take a different
starting point, as described in later sections.

The wavefunction of the paired quantum Hall state 
constructed in this way can be written as
\begin{eqnarray}
\Psi_{\rm pair} = 
\Psi^{\rm cf}_{\rm pair}
\prod_{i>j} (z^{\uparrow}_i-z^{\uparrow}_j)^2
\prod_{k>l} (z^{\downarrow}_k-z^{\downarrow}_l)^2 \ ,
\end{eqnarray} 
where 
\begin{eqnarray}
\Psi^{\rm cf}_{\rm pair}
&=& {\rm Pf} [f(z_i, z_j ; \uparrow, \downarrow)] \cr
&\equiv& {\cal A} [ f(z_1, z_2 ; \uparrow, \downarrow)
f(z_3, z_4; \uparrow, \downarrow) \cdots ] \ .   
\end{eqnarray}
and$f(z_1, z_2 ; \uparrow, \downarrow)$ is the pair 
wavefunction which depends on the symmetry of the
pairing order parameter. ${\rm Pf} [\cdots]$ denotes the
Pfaffian, which is defined in the second line,
with ${\cal A} [\cdots]$ denoting the antisymmetrized
product.  
Notice that $\Psi_{\rm pair}$ can be
regarded as the product of the wavefunction of
the paired composite fermions and that of 
the (2,2,0) bosonic Laughlin quantum Hall state.

It is not immediately clear what choice
of $f(z_1, z_2 ; \uparrow, \downarrow)$ 
is most favorable energetically.
In the Chern-Simons theory of \cite{bonesteel},
there is a pairing instability in all angular
momentum channels \cite{bonesteel}.
In a modified Chern-Simons theory, it was
claimed that the leading instability occurs in the $p$-wave
channel \cite{morinari}. These approximate
calculations do not necessairly capture
the detailed energetics which determines
the pairing symmetry. Hence, we will not
enter into a discussion of energetics, but
limit ourselves to a discussion of the
simplest (and, therefore, likeliest)
possibilities.

The simplest possibility is $p_x + i p_y$ pairing,
\begin{equation}
\Psi_{\rm pair} = {\rm Pf} \left [ {\uparrow_i \downarrow_j + 
\downarrow_i \uparrow_j \over z_i - z_j } \right ]
\prod_{i>j} (z^{\uparrow}_i-z^{\uparrow}_j)^2
\prod_{k>l} (z^{\downarrow}_k-z^{\downarrow}_l)^2 \ .
\end{equation} 
Using the Cauchy identity \cite{ho},
\begin{equation}
\label{eqn:Cauchy}
\prod^N_{i>j=1}(a_i-a_j)(b_i-b_j)=\prod^N_{i,j=1} (a_i-b_j)
{\rm Det}|(a_i-b_j)^{-1}| \ ,
\end{equation}  
this can be rewritten as
\begin{eqnarray}
\Psi_{\rm pair} &=& \Psi_{(3,3,-1)} \cr
&=& \prod_{i>j} (z^{\uparrow}_i-z^{\uparrow}_j)^3
(z^{\downarrow}_i-z^{\downarrow}_j)^3
\prod_{i,j} (z^{\uparrow}_i-z^{\downarrow}_j)^{-1} \ .
\end{eqnarray} 
Thus $\Psi_{\rm pair}$ is the $(3,3,-1)$
state if one takes the $p_x+ip_y$ pairing. 
This wavefunction is well-behaved
in the long distance
limit, but has a short distance singularity.
In the presence of Landau-level mixing, the short distance 
part of the wavefunction can be modified without changing
the structure of the wavefunction in the long distance limit.
\begin{eqnarray}
\label{eqn:(3,3,-1)}
\Psi_{\rm pair} &=& {\rm Pf} \left [
f\left(\left|{z_i} - {z_j}\right|/\xi\right)\:
{\uparrow_i \downarrow_j + 
\downarrow_i \uparrow_j \over z_i - z_j } \right ]\,\times
\cr & &{\hskip 1 cm}
\prod_{i>j} (z^{\uparrow}_i-z^{\uparrow}_j)^2
\prod_{k>l} (z^{\downarrow}_k-z^{\downarrow}_l)^2 \ .
\end{eqnarray}
Here, $f(0)=0$ and $f(x)\rightarrow 1$ as $x\rightarrow \infty$.
In realistic systems, where Landau-level mixing
is substantial, $\Psi_{(3,3,-1)}$ could be
a good candidate for the paired quantum Hall state represented 
by $\Psi_{\rm pair}$.
It is natural to assume that $\Psi_{\rm pair}$
does not have an ``interlayer Josephson effect'' because
there is no gapless neutral mode in the system, in contrast
to the case of the $(1,1,1)$ state \cite{eisenstein,demler}.
We will show this later by direct calculation.

Another possibility for the pair wavefunction 
$f (z_i, z_j; \uparrow, \downarrow)$ is the exponentially
decaying function
\begin{equation}
\label{eqn:other_pairing}
f (z_i, z_j; \uparrow, \downarrow) =
\left({\uparrow_i \downarrow_j + 
\downarrow_i \uparrow_j }\right)
(z_i-z_j) e^{-|z_i-z_j|/\xi}
\end{equation}
with a correlation
length $\xi$. This would correspond to a ``strong'' pairing
state while the previous choice of $1/(z_i-z_j)$ corresponds
to a ``weak'' pairing state in the terminology of
of Read and Green \cite{green}. Finally, the
pairing wavefunction could have $s$-wave symmetry, which
would entail a pairing form similar to 
(\ref{eqn:other_pairing}), but without the
$(z_i-z_j)$ factor.

\section{$\lowercase{d}\sim \ell_B$: $(1,1,1)$ state}

When the layer separation becomes sufficiently small,
the interlayer Coulomb interaction can be comparable 
to or even larger than the intralayer
Coulomb interaction. 
In this case, it should be more advantageous to
first form an ``interlayer'' dipolar object which consists of 
an electron in one layer and two vortices in the other
layer, then form a paired state of these ``interlayer''
composite fermions, as shown in Fig.3. 
In the Chern-Simons formulation, this corresponds
to the situation in which the electron in one layer can
only see fictitious flux in the other layer.
The appropriate Chern-Simons constraint equation are:
\begin{equation}
\nabla \times {\bf a}_{\uparrow} = 2 \pi {\tilde \phi} 
\delta \rho_{\downarrow} \ ,   \ \ \
\nabla \times {\bf a}_{\downarrow} = 2 \pi {\tilde \phi} 
\delta \rho_{\uparrow} \ .
\end{equation}

As in the previous case, we can form symmetric and antisymmetric
combinations of the gauge fields ${\bf a}_{\uparrow}$
and ${\bf a}_{\downarrow}$. Again, the antisymmetric
combination mediates an attractive interaction.
The wavefunction of the corresponding paired quantum Hall state
has the following form.
\begin{equation}
\Phi_{\rm pair} = 
\Phi^{\rm cf}_{\rm pair}
\prod^N_{i,j=1} (z^{\uparrow}_i-z^{\downarrow}_j)^2 \ ,
\end{equation} 
where 
\begin{equation}
\Phi^{\rm cf}_{\rm pair}
= {\rm Pf} [g(z_i, z_j ; \uparrow, \downarrow)]
\end{equation}
and $g(z_1, z_2 ; \uparrow, \downarrow)$ is the appropriate
pair wavefunction.
Notice that the wavefunction $\Phi_{\rm pair}$
can be regarded as the product of the wavefunction of
a paired state of the composite fermions and that of
the (0,0,2) bosonic Laughlin quantum Hall state.

\begin{figure}    
\center    
\centerline{\epsfysize=1.4in    
\epsfbox{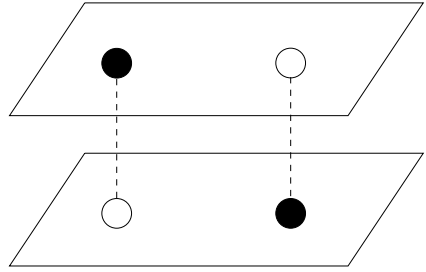}}
\begin{minipage}[t]{8.1cm}    
\caption{The pairing of the ``interlayer''
composite fermions.
}  
\label{fig3}   
\end{minipage}    
\end{figure}    

However, this line of thinking appears to
conflict with the conventional wisdom
that a bi-layer quantum Hall system
at $\nu_T=1$ is described by the $(1,1,1)$
state for $d/{\ell_B}\sim 1$ \cite{Halperin}.
Fortunately, the $(1,1,1)$ state,
\begin{equation}
\Psi_{(1,1,1)} = \prod^{N}_{i>j=1} (z^{\uparrow}_i-z^{\uparrow}_j)
(z^{\downarrow}_i-z^{\downarrow}_j)
\prod^{N}_{i,j=1} (z^{\uparrow}_i-z^{\downarrow}_j) \ .
\end{equation}
can be rewritten in the form
\begin{equation}
\Psi_{(1,1,1)} = {\rm Pf} \left [ {\uparrow_i \downarrow_j + 
\downarrow_i \uparrow_j \over z_i - z_j } \right ]
\prod^{N}_{i,j=1} (z^{\uparrow}_i-z^{\downarrow}_j)^2 \ .
\end{equation}
using the Cauchy identity.
In other words,
\begin{equation}
g (z_i, z_j ; \uparrow, \downarrow) = 
{\uparrow_i \downarrow_j + 
\downarrow_i \uparrow_j \over z_i - z_j } \ . 
\end{equation}
is the correct choice for $d/{\ell_B}\sim 1$.

Notice that this form of $g (z_i, z_j ; \uparrow, \downarrow)$
corresponds to the (pseudo-)spin triplet $p_x + i p_y$ pairing 
order parameter for the ``interlayer'' composite fermions. 
From this point of view, it is natural to have the same
pairing symmetry (\ref{eqn:(3,3,-1)}) for $d>\ell_B$ 
but for ``intralayer'' rather
than ``interlayer'' composite fermions.

\section{Drag Resistivities}

In the preceeding sections, we have described
three possible states of a
bi-layer system at $\nu_T=1$. As one decreases the ratio
of $d/\ell_B$, we expect the following sequence
of phase transitions
\begin{equation}
\Psi_{\rm compressible} \rightarrow \Psi_{(3,3,-1)}
\rightarrow 
\Psi_{(1,1,1)} \ .
\end{equation}
The transition between $\Psi_{\rm compressible}$ and
$\Psi_{(3,3,-1)}$ is disorder-driven since
$\Psi_{\rm compressible}$ should never be the ground
state in the presence of disorder. Hence, we
expect the transition to be second-order.
On the other hand, the transition between
$\Psi_{(3,3,-1)}$ and $\Psi_{(1,1,1)}$
is likely to be first order
because the nature of the underlying quasiparticles changes 
from the ``intralayer'' composite fermions to the ``interlayer''
composite fermions and this change is not likely to be
continuous. More formally, the change of the wavefunction
requires a flux attachment transformation without changing 
the total filling factor which does not usually occur 
continuously. Of course, disorder will eventually
round the transition and make it second order.

Since the first order transition may be obscured
by disorder, it is important to discuss experiments 
which can distinguish the
$\Psi_{(3,3,-1)}$ and $\Psi_{(1,1,1)}$ states.
Here, we propose Coulomb drag experiments in which
the longitudinal, $\rho^{\uparrow \downarrow}_{xx}$ and 
Hall, $\rho^{\uparrow \downarrow}_{xy}$, drag resistivities 
are used to distinguish the different 
phases. These may be calculated in Chern-Simons
theory from
\begin{equation}
\label{eqn:drag}
{\rho_{ij}^{\alpha\beta}} = {\rho_{ij}^{{\rm cf}\:\alpha\beta}}
+ {\epsilon_{ij}}\: {\rho^{{\rm cs}\:\alpha\beta}}
\end{equation}
where $i,j=x,y$, $\alpha,\beta=\uparrow,\downarrow$,
$\epsilon_{ij}$ is the antisymmetric tensor,
and ${\rho^{{\rm cs}\:\alpha\beta}}=2{\delta_{\alpha\beta}}$
in the compressible and $(3,3,-1)$ states
(``intralayer'' composite fermions)
while ${\rho^{{\rm cs}\:\alpha\beta}}=2{\sigma^x_{\alpha\beta}}$
in the $(1,1,1)$ state (``interlayer'' composite fermions).
 
First, consider the longitudinal drag resistivity.
In the compressible state, if we neglect gauge field
fluctuations, $\rho_{xx}^{{\rm cf}\:\uparrow \downarrow}=0$,
so $\rho_{xx}^{\uparrow \downarrow}=0$. Including
these fluctuations,
it vanishes as $T^{4/3}$ at low temperatures \cite{kim}.
In $\Psi_{(1,1,1)}$ and $\Psi_{(3,3,-1)}$,
$\rho_{xx}^{{\rm cf}\:\uparrow \downarrow}$ and, hence,
$\rho_{xx}^{\uparrow \downarrow}$
vanish at zero temperature and are activated
at low temperatures.

Now, let us consider the Hall drag resistivity.
In the compressible state, both terms on
the right-hand-side of (\ref{eqn:drag})
vanish, so the Hall drag resistivity vanishes.
In the $(3,3,-1)$ state, ${\rho^{\rm cf}_{xy}}$
is that of a $p_x + i p_y$ superconductor,
which has vanishing charge resistivity
(since it is a superconductor) but quantized
spin Hall resistivity. In other words,
${\rho_{xy}^{{\rm cf}\: cc}}=0$,
${\rho_{xy}^{{\rm cf}\: cs}}=0$,
${\rho_{xy}^{{\rm cf}\: ss}}=1$.
Consequently, ${\rho_{xy}^{\uparrow \uparrow}}=
{\rho_{xy}^{\downarrow \downarrow}}=3$,
and ${\rho_{xy}^{\uparrow \downarrow}}=-1$.
In the $(1,1,1)$ state, ${\rho^{\rm cf}_{xy}}$
is identical, but ${\rho^{\rm cs}_{xy}}$
is different, so ${\rho_{xy}^{\uparrow \uparrow}}=
{\rho_{xy}^{\downarrow \downarrow}}=1$,
and ${\rho_{xy}^{\uparrow \downarrow}}=1$.

We conclude by summarizing our results. We
have shown that the $\nu={1\over 2}$ (${\nu_T}=1$)
bilayer quantum Hall system (in the absence of interlayer
tunneling) is likely to have as its ground state a
novel paired Hall state (possibly of $p$-wave symmetry)
for intermediate layer separations $d>{\ell_B}$,
which gives way to the usual $(1,1,1)$ state
for smaller layer separations ($d\sim {\ell_B}$),
and to compressible Fermi-liquid-type states (two
decoupled Halperin-Lee-Read \cite{HLR}
$\nu={1\over 2}$ layers) for large layer separations
($d\gg{\ell_B}$). We argue that the quantum phase transitions
separating the paired states from the $(1,1,1)$ and
bilayer Halperin-Lee-Read states can be experimentally
studied via the measurement of various components
of interlayer drag resistivities.

\begin{acknowledgements}
We would like to thank L. Balents
and, especially, N. Read for discussion.
YBK, CN, and SDS would like to thank
the Aspen Center for Physics for hospitality.
This work was supported by the NSF
under grant numbers DMR-9983783 (YBK)
and DMR-9983544 (CN), the A.P. Sloan Foundation (YBK and CN),
the Harvard Society of Fellows (ED), and the
ONR (SDS).
\end{acknowledgements}

\end{multicols}


\begin{references}


\bibitem{footnote_1}
Throughout this paper, we use $\nu_T$ to denote the total
filling factor and $\nu$ to denote the filling
factor in each layer, ${\nu_T}=2\nu$.


\bibitem{Eisenstein97}
See, for example, the articles by J.~P. Eisenstein,
S.~M. Girvin and A.~H. MacDonald in {\it Perspectives
in Quantum Hall Effects}, editied by S. Das Sarma
and A. Pinczuk (Wiley, New York, 1997); and references therein.


\bibitem{Halperin} B. I. Halperin, Helv. Phys. Acta {\bf 56}, 75 (1983);
Surf. Sci. {\bf 305},1 (1994).


\bibitem{eisenstein} 
I. B. Spielman, J. P. Eisenstein, L. N. Pfeiffer, and K. W. West,
cond-mat/0002387.


\bibitem{Wen92}
X.~G. Wen and A. Zee, Phys. Rev. Lett.
{\bf 69}, 1811 (1992); Pys. Rev. B {\bf 47},
2265 (1993).

\bibitem{Moore91}
G. Moore and N. Read, Nucl. Phys.
B {\bf 360}, 362 (1991).

\bibitem{bonesteel} N. E. Bonesteel, Phys. Rev. B {\bf 48}, 11484 (1993);
N. E. Bonesteel, I. A. MacDonald, and C. Nayak,
Phys. Rev. Lett. {\bf 77}, 3009 (1996);
M. Greiter, F. Wilczek, and X.-G. Wen, Phys. Rev. Lett.
{\bf 66}, 3205 (1991). 

\bibitem{demler}
E. Demler, C. Nayak, and S. Das Sarma, cond-mat/0008137. 

\bibitem{jain}
J. K. Jain, Phys. Rev. Lett. {\bf 63}, 199 (1989);
Adv. Phys. {\bf 41}, 105 (1992).

\bibitem{rezayi}
E. Rezayi and N. Read, Phys. Rev. Lett. {\bf 72}, 
900 (1994); {\it ibide}. {\bf 73}, 1052 (1994).

\bibitem{girvin}
S. M. Girvin and T. Jach, 
Phys. Rev. B {\bf 29}, 5617 (1984).

\bibitem{CS}
A. Lopez and E. Fradkin, 
Phys. Rev. B {\bf 44}, 5246 (1991);
V. Kalmeyer and S.-C. Zhang, Phys. Rev. B {\bf 46}, 
9889 (1992).

\bibitem{HLR}
B. I. Halperin, P. A. Lee, and N. Read, 
Phys. Rev. B {\bf 47}, 7312 (1993).

\bibitem{read94}
N. Read, Semi. Sci. Tech. {\bf 9}, 1859 (1994); 
Surf. Sci. {\bf 361}, 7 (1996).

\bibitem{shankar} R. Shankar and G. Murthy, 
Phys. Rev. Lett. {\bf 79}, 4437 (1997);
D. H. Lee, Phys. Rev. Lett. {\bf 80}, 4745 (1998);
V. Pasquier and F. D. M. Haldane,
Nucl. Phys. {\bf B 516}, 719 (1998).

\bibitem{stern} B. I. Halperin and A. Stern, Phys. Rev. Lett. 
{\bf 80}, 5457 (1998); A. Stern {\it et al.}, 
Phys. Rev B {\bf 59}, 12547 (1999).

\bibitem{read98} N. Read, Phys. Rev. B {\bf 58}, 16262 (1998).


\bibitem{ho} T. L. Ho, Phys. Rev. Lett. {\bf 75}, 1186 (1995).

\bibitem{morinari} T. Morinari, Phys. Rev. Lett. {\bf 81}, 3741 (1998).


\bibitem{green}
N. Read and D. Green, 
Phys. Rev. B {\bf 61}, 10267 (2000).

\bibitem{kim} Y. B. Kim and A. J. Millis, 
Physica E {\bf 4}, 171 (1999) and cond-mat/9611125;
I. Ussishkin and A. Stern, Phys. Rev. B {\bf 56}, 4013 (1997);
S. Sakhi, Phys. Rev. B {\bf 56}, 4098 (1997).

\bibitem{yang} K. Yang, Phys. Rev. B {\bf 58}, R4246 (1998).

\bibitem{vignale} G. Vignale and A. H. MacDonald, Phys. Rev. Lett.
{\bf 76}, 2786 (1996);  
I. Ussishkin and A. Stern, Phys. Rev. Lett. {\bf 81}, 3932 (1998);
F. Zhou and Y. B. Kim, 
Phys. Rev. B {\bf 59}, R7825 (1999).
  
\bibitem{senthil} T. Senthil and M. P. A. Fisher, Phys. Rev. B
{\bf 60}, 4245 (1999).

\end{references}
\end{document}